# Spectroscopic and Photometric Study of the S-type Mira V667 Cassiopeiae and the Carbon Star OR Cephei

**David Boyd**

*West Challow Observatory, OX12 9TX, UK; davidboyd@orion.me.uk*

**Abstract**　We report on spectroscopic and photometric observations of the S-type Mira V667 Cas and the carbon star Mira OR Cep recorded during one pulsation cycle of each star in 2022. Spectra are calibrated in absolute flux using concurrent photometry. We present measurements of V and $I_c$ magnitudes and V–$I_c$ color index, classify spectra in the revised MK system, and investigate how absolute flux in the Hα to Hδ emission lines varies with pulsation phase for each star.

## 1. Introduction

### 1.1. M-type giants

Towards the end of their lives, M-type giant stars are increasing in size and luminosity, ascending the Asymptotic Giant Branch (AGB) of the Hertzsprung-Russell Diagram, and losing mass by a stellar wind. Convection brings carbon, oxygen, and other products from nuclear burning, plus heavier s-process elements formed by slow neutron capture, into the upper layers of the star by a process called dredge-up. The proportion of these different elements in the outer layers of the giant star determines its chemical nature. This is often characterized in terms of a carbon/oxygen ratio (C/O).

• If there is more oxygen than carbon in the outer layer of the star (C/O < 1), it is described as an oxygen-rich M giant. Molecular oxides of titanium and other metals form in the star's atmosphere and absorb light at specific wavelengths, creating a characteristic saw-tooth spectral profile.

• If oxygen and carbon are present in approximately equal amounts (C/O ~ 1), the star is referred to as an S-type or S star. These stars have absorption lines and bands in their spectra due to the presence of s-process elements, with bands of zirconium oxide being the signature of S stars.

• If carbon predominates (C/O > 1), it is called a carbon star and the atmosphere contains compounds of carbon such as carbon monoxide (CO), diatomic carbon (C2), cyanogen (CN), and CH. Carbon in the star's atmosphere tends to hide the absorption features caused by other molecules.

At this late stage in its life, the giant has a very swollen convective envelope and atmosphere about 1 AU in radius. These are eventually driven off by a strongly enhanced wind, leaving a hot white dwarf consisting mostly of carbon and oxygen. During this post-AGB stage, some giants may form planetary nebulae through processes which are currently not well understood.

### 1.2. Pulsation in AGB stars

Stars on the AGB often experience pulsations with a period of the order of a year driven by the competition between radiation pressure and gravity. These pulsations accelerate the loss of material from the star's atmosphere. The process driving pulsation is called the kappa-mechanism and its physical causes are still being debated (Querci 1986; Fleischer *et al*. 1995; Höfner *et al*. 1995; Smith *et al*. 2002). Pulsation generates radial shocks in the stellar atmosphere (Gorbatskii 1961) which ionize hydrogen and produce emission lines of the Balmer series, see for example Willson (1976), Fox *et al*. (1984), Gillet (1988). As these lines are produced relatively low down in the stars' atmospheres, their flux is partially absorbed by molecules higher in the atmosphere.

## 2. Observations

We have previously reported on the spectroscopic and photometric behavior of the oxygen-rich Mira stars SU Cam and RY Cep (Boyd 2021) and T Cep (Boyd 2023). Here we report similar observations and analysis of the S-type Mira V667 Cas, an omicron Ceti-type variable, and the carbon star Mira OR Cep. As the spectral profile of both stars is strongly peaked towards the red, we decided to use V and $I_c$ photometric filters. We recorded V and $I_c$ band photometry and low resolution (R ~ 1100) spectra of both stars with an approximately monthly cadence during 2022. Equipment used and data reduction method are similar to those described in Boyd (2021, 2023).

Photometry was obtained with a 0.35-m SCT operating at f/5 and equipped with Astrodon photometric filters and a temperature-regulated SXVR-H9 CCD camera. Photometric observations were made alternatively through V and $I_c$ filters, with typically 10 images recorded in each filter. V magnitudes of five nearby comparison stars in the field of each star were obtained from the AAVSO Photometric All-Sky Survey (APASS; Henden *et al*. 2018). $I_c$ magnitudes for these stars were derived from their Sloan ugriz magnitudes in APASS using transformations in Jester *et al*. (2005). These photometric observations were made concurrently while recording spectra. All photometric images were bias, dark, and flat corrected and instrumental magnitudes obtained by aperture photometry using the software AIP4WIN (Berry and Burnell 2005). Instrumental V and $I_c$ magnitudes were transformed to the Johnson-Cousins photometric standard using V–$I_c$ color indices and color transformation coefficients, airmass, and first order extinction coefficients, and zero points for each filter as described in (Boyd 2012). Times were recorded as Julian Date (JD) but no barycentric correction was applied as their radial velocities are expected to be essentially constant.

The large V–$I_c$ color index of both stars (between 3.1 and 5.6) poses a problem. The normal process of determining color transformation coefficients uses fields of standard stars but these generally do not have such large V–$I_c$ color indices. We have therefore had to extrapolate these coefficients beyond the region in which they have been measured. We take this into account by increasing the uncertainty in the coefficient by a factor of 4 in the case of V667 Cas and a factor of 3 in the case of OR Cep and propagating this into the uncertainty in the derived magnitudes.

We used V-band magnitudes of both stars in the AAVSO international Database (AAVSO AID; Kafka 2022) going back four years, including our own measurements, to establish their epochs of maximum light in 2022 (pulsation phase 0) and pulsation periods for use in this study. These are listed in Table 1.

The Julian Date, pulsation phase, nightly mean V and $I_c$ magnitudes, and V–$I_c$ color index for our measurements of V667 Cas during the 2022 pulsation cycle are listed in Table 2. Average uncertainties in V and $I_c$ are ±0.03 and in V–$I_c$ is ±0.045. Nightly mean V and $I_c$ magnitudes and V–$I_c$ color indices for V667 Cas, phased according to the parameters in Table 1, are shown in Figure 1.

The Julian Date, pulsation phase, nightly mean V and $I_c$ magnitudes, and V–$I_c$ color index for our measurements of OR Cep during the 2022 pulsation cycle are listed in Table 3. Average uncertainty in V is ±0.02, in $I_c$ is ±0.03, and in V–$I_c$ is ±0.04. Nightly mean V and $I_c$ magnitudes and V–$I_c$ color indices for OR Cep, phased according to the parameters in Table 1, are shown in Figure 2.

Figure 3 shows how the V–$I_c$ color index varies with V magnitude for both V667 Cas and OR Cep. In both stars the relationship is close to linear, with the star becoming bluer as it brightens and redder as it fades. There is a small hysteresis effect. The initial point in each cycle is marked with a larger dot.

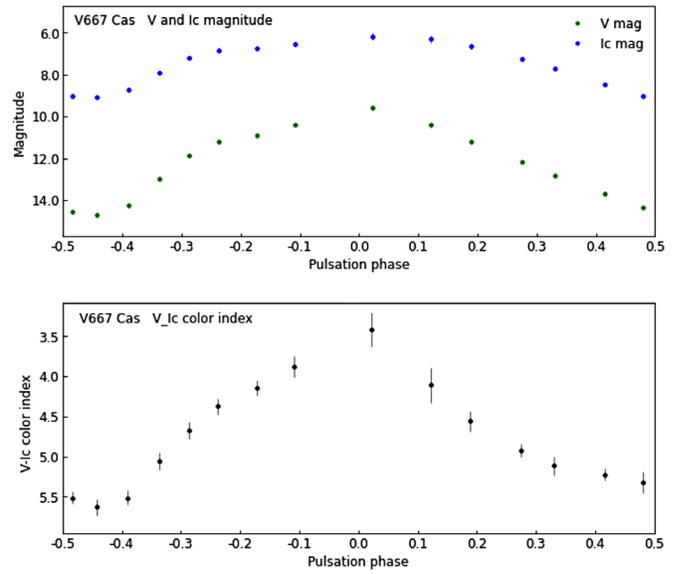

Figure 1. Variation of V and $I_c$ magnitudes and V–$I_c$ color index with pulsation phase for V667 Cas during the 2022 pulsation cycle. If not visible, errors on V and $I_c$ magnitudes are within the symbol plotted.

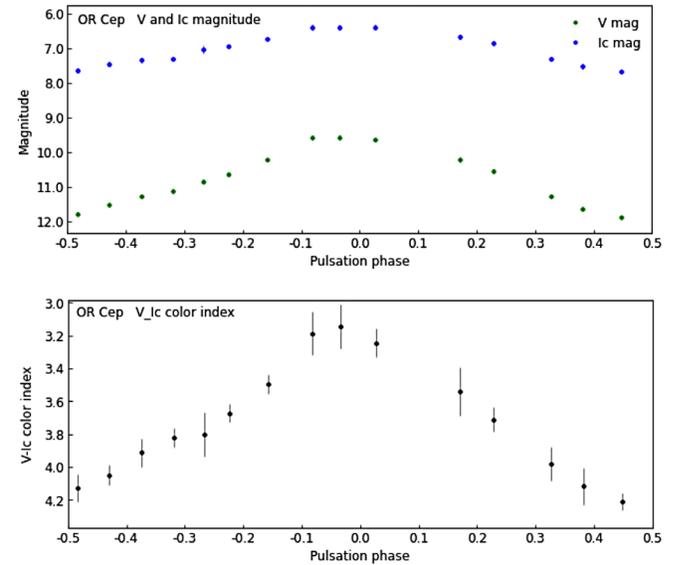

Figure 2. Variation of V and $I_c$ magnitudes and V–$I_c$ color index with pulsation phase for OR Cep during the 2022 pulsation cycle. If not visible, errors on V and $I_c$ magnitudes are within the symbol plotted.

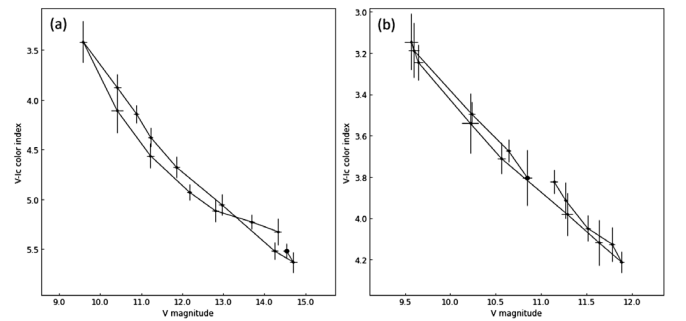

Figure 3. Variation of V–$I_c$ color index with V magnitude in (a) V667 Cas and (b) OR Cep. The initial point in each cycle is marked with a larger dot.

Table 1. Epochs of maximum light (pulsation phase 0) and pulsation periods for V667 Cas and OR Cep used in this study.

| Star | Epoch of maximum light (JD) | Current pulsation period (d) |
|---|---|---|
| V667 Cas | 2459758 | 340 |
| OR Cep | 2459685 | 347 |

Table 2. Julian Date, pulsation phase, V and $I_c$ magnitudes and V–$I_c$ color index for V667 Cas during the 2022 pulsation cycle.

| Julian Date | Phase | V (mag) | $I_c$ (mag) | V–$I_c$ (mag) |
|---|---|---|---|---|
| 2459593 | –0.48 | 14.54 | 9.02 | 5.52 |
| 2459607 | –0.44 | 14.71 | 9.08 | 5.63 |
| 2459625 | –0.39 | 14.25 | 8.73 | 5.52 |
| 2459643 | –0.34 | 12.97 | 7.91 | 5.05 |
| 2459660 | –0.29 | 11.86 | 7.18 | 4.68 |
| 2459677 | –0.24 | 11.23 | 6.85 | 4.38 |
| 2459699 | –0.17 | 10.88 | 6.74 | 4.14 |
| 2459721 | –0.11 | 10.42 | 6.54 | 3.88 |
| 2459753 | –0.01 | 9.46 | 6.16 | 3.30 |
| 2459765 | 0.02 | 9.59 | 6.17 | 3.42 |
| 2459799 | 0.12 | 10.42 | 6.31 | 4.11 |
| 2459821 | 0.19 | 11.17 | 6.60 | 4.56 |
| 2459822 | 0.19 | 11.22 | 6.66 | 4.56 |
| 2459851 | 0.27 | 12.17 | 7.25 | 4.93 |
| 2459870 | 0.33 | 12.80 | 7.69 | 5.12 |
| 2459899 | 0.42 | 13.69 | 8.46 | 5.23 |
| 2459921 | 0.48 | 14.33 | 9.00 | 5.33 |

*Notes: Average uncertainties in V and $I_c$ are ±0.03 and in V–$I_c$ is ±0.045.*

Table 3. Julian Date, pulsation phase, V and $I_c$ magnitudes and V–$I_c$ color index for OP Cep during the 2022 pulsation cycle.

| Julian Date | Phase | V (mag) | $I_c$ (mag) | V–$I_c$ (mag) |
|---|---|---|---|---|
| 2459592 | –0.27 | 10.85 | 7.04 | 3.80 |
| 2459607 | –0.22 | 10.64 | 6.97 | 3.67 |
| 2459630 | –0.16 | 10.23 | 6.74 | 3.50 |
| 2459656 | –0.08 | 9.60 | 6.42 | 3.19 |
| 2459673 | –0.03 | 9.57 | 6.42 | 3.14 |
| 2459694 | 0.03 | 9.65 | 6.40 | 3.24 |
| 2459744 | 0.17 | 10.22 | 6.68 | 3.54 |
| 2459764 | 0.23 | 10.56 | 6.85 | 3.71 |
| 2459798 | 0.33 | 11.29 | 7.31 | 3.98 |
| 2459817 | 0.38 | 11.63 | 7.52 | 4.12 |
| 2459840 | 0.45 | 11.88 | 7.67 | 4.21 |
| 2459864 | –0.48 | 11.78 | 7.65 | 4.13 |
| 2459883 | –0.43 | 11.51 | 7.46 | 4.05 |
| 2459902 | –0.37 | 11.27 | 7.36 | 3.91 |
| 2459921 | –0.32 | 11.14 | 7.32 | 3.82 |

Notes: Average uncertainty in V is ±0.02, in $I_c$ is ±0.03 and in V–$I_c$ is ±0.04.

Spectroscopy was obtained with a 0.28-m Schmidt-Cassegrain Telescope (SCT) operating at f/5 and equipped with an auto-guided Shelyak LISA slit spectrograph and a SXVR-H694 CCD camera. The slit width was 23μ, giving a mean spectral resolving power of ~1000. Spectra were processed with the ISIS spectral analysis software (ISIS; Buil 2021). Spectroscopic images were bias, dark, and flat corrected, geometrically corrected, sky background subtracted, spectrum extracted, and wavelength calibrated using the integrated ArNe calibration source. They were then corrected for instrumental and atmospheric losses using spectra of a nearby star with a known spectral profile from the MILES library of stellar spectra (Falcón-Barroso et al. 2011) situated as close as possible in airmass to the target star and obtained immediately prior to the target spectra. Typically, 12 five-minute guided spectral integrations were recorded for each star and combined into the final spectrum, giving signal-to-noise ratios ranging from ~100 at maximum brightness to ~10 at minimum. Spectra were calibrated in absolute flux in FLAM units as erg/cm²/s/Å using V magnitudes measured concurrently with the spectra as described in Boyd (2020).

From Gaia DR3, the distance to V667 Cas is 1159 +60 –55 parsecs and to OR Cep is 1036 +32 –28 parsecs (Bailer-Jones et al. 2021). According to Schlafly and Finkbeiner (2011), the total galactic reddening in the direction of V667 Cas is E(B–V) = 0.55 and in the case of OR Cep is E(B–V) = 0.90. The galactic coordinates for V667 Cas are l=135°, b=+11.4° and for OR Cep are l=119°, b=+7.5°. Based on these coordinates and distances, V667 Cas is placed at 229 parsecs and OR Cep 135 parsecs above the plane of the galaxy, so they will have experienced a large proportion of this reddening. We have therefore dereddened our spectra of both stars using these color excesses assuming $R_v$, the average galactic ratio of total to selective reddening in the visual band, is 3.1, and using the formulae in Cardelli et al. (1989). Magnitudes measured photometrically have not been dereddened.

All measured magnitudes and spectra are available in the British Astronomical Association (BAA) Photometry Database (BAA Photometry Database 2021) and the BAA Spectroscopy Database (BAA Spectroscopy Database 2021), respectively. Nightly mean magnitudes are available in the AAVSO International Database (Kafka 2022). We developed Python software for our analysis which made extensive use of the Astropy package (Astropy Collaboration et al. 2018).

### 3. Spectral classification

3.1. Classifying our V667 Cas spectra

Figure 4 shows all our dereddened and absolute flux calibrated spectra of V667 Cas recorded during the 2022 pulsation cycle marking the positions of the first 4 Balmer emission lines, ZrO molecular band heads, and Na I D absorption lines.

Keenan and Boeshaar (1980) give spectral classification criteria on the revised MK system for S-type stars based on a C/O index which ranges from SX/1 to SX/7, where X is a temperature parameter relating to the temperature sequence of M-giant stars and the number is an indicator of the relative strengths of TiO and ZrO molecular bands and the strength of Na I D absorption lines. Note that this C/O index is not directly related to the C/O ratio mentioned above; see also Gray and Corbally (2009).

Our spectra of V667 Cas in Figure 4 show strong Na I D absorption lines and ZrO molecular band heads but no visible presence of TiO bands. According to the Keenan and Boeshaar criteria, this indicates the spectral type of our V667 Cas spectra is probably either SX/6 or SX/7. We tried to measure the 6450/6456 Å ratio as a temperature criterion but these features are indistinct in our spectra.

We therefore decided to classify our spectra of V667 Cas by comparing their spectral features with spectra in Table 2 of Keenan and Boeshaar. The most consistent match appeared to be with R Gem, which according to Keenan and Boeshaar has a spectral type SX/6 where X ranges from 3.5 to 6.5. This suggests the spectral type of V667 Cas during this pulsation cycle was most probably SX/6. Figure 5 compares a spectrum of R Gem from the AAVSO Spectroscopic Database (AAVSO AVSpec) recorded by BRAJ on 2023 March 1 using an ALPY spectrograph (R~600) with one of our spectra of V667 Cas, both spectra having been recorded close to the peak of a pulsation cycle. Their similarity supports our classification.

3.2. Classifying our OR Cep spectra

Figure 6 shows our dereddened and absolute flux-calibrated spectra of OR Cep with Balmer lines, C2 and CN molecular band heads, and Na I D lines marked.

As noted in Gray and Corbally (2009), the carbon stars encompass a wide variety of objects of different populations and origins. The relative abundance of carbon over oxygen in these stars means that almost all the oxygen is taken up in the formation of CO, leaving little to form molecules such as TiO or ZrO. Spectra are dominated by molecular bands of the carbon-based molecules C2, CN, and CH. In his Revised MK Spectral Classification of the Red Carbon Stars, Keenan (1993) introduced a classification system based on five types of carbon stars. The strengths of the C2 Swan bands close to 5000 Å are normally used to assign a spectral type, but in our low-resolution spectra these are too indistinct to be useful. Our low flux below 4750 Å also means that we are unable to use the Ba II 4554/Sr I 4607 Å line ratio as a temperature criterion.

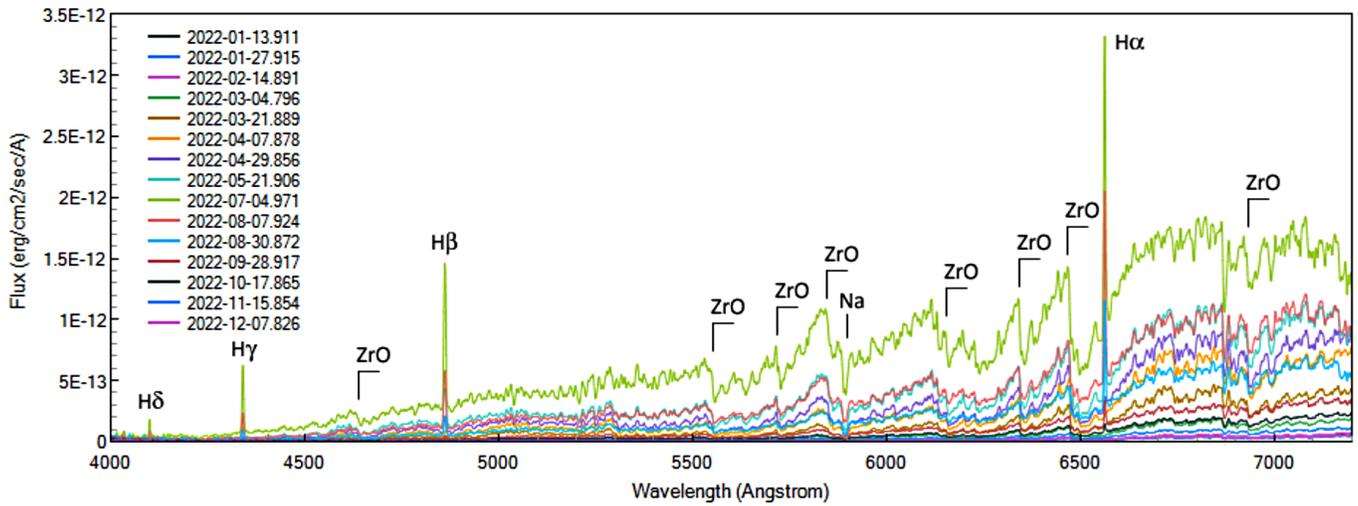

Figure 4. All flux calibrated spectra of V667 Cas during the 2022 pulsation cycle marking the positions of the first 4 Balmer emission lines, the ZrO molecular band heads, and the Na D absorption lines.

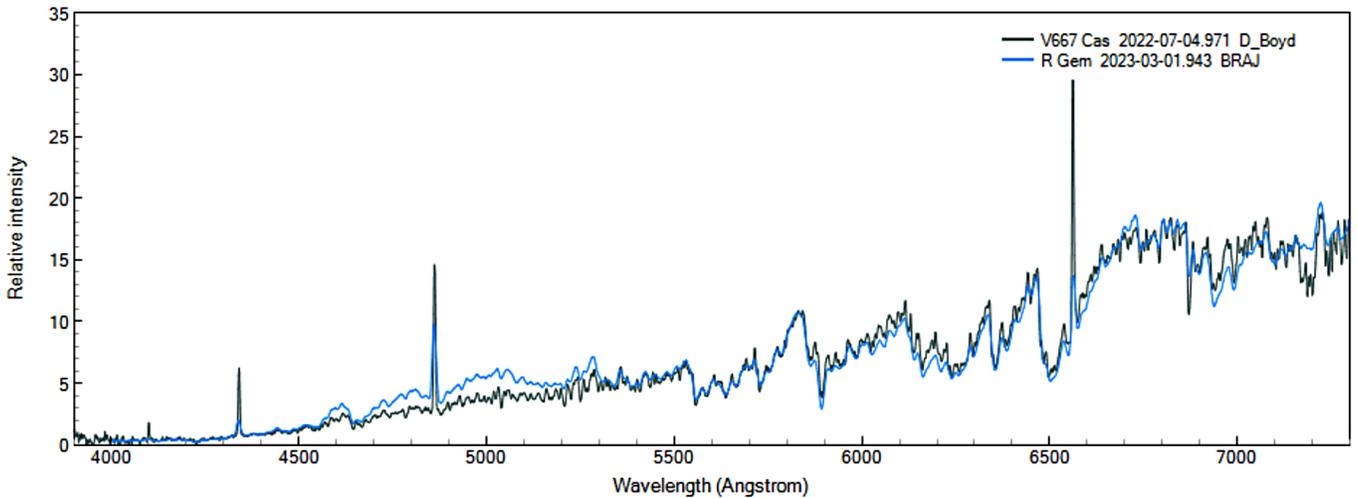

Figure 5. Comparison of spectra of V667 Cas and R Gem taken at a peak of their pulsation cycles.

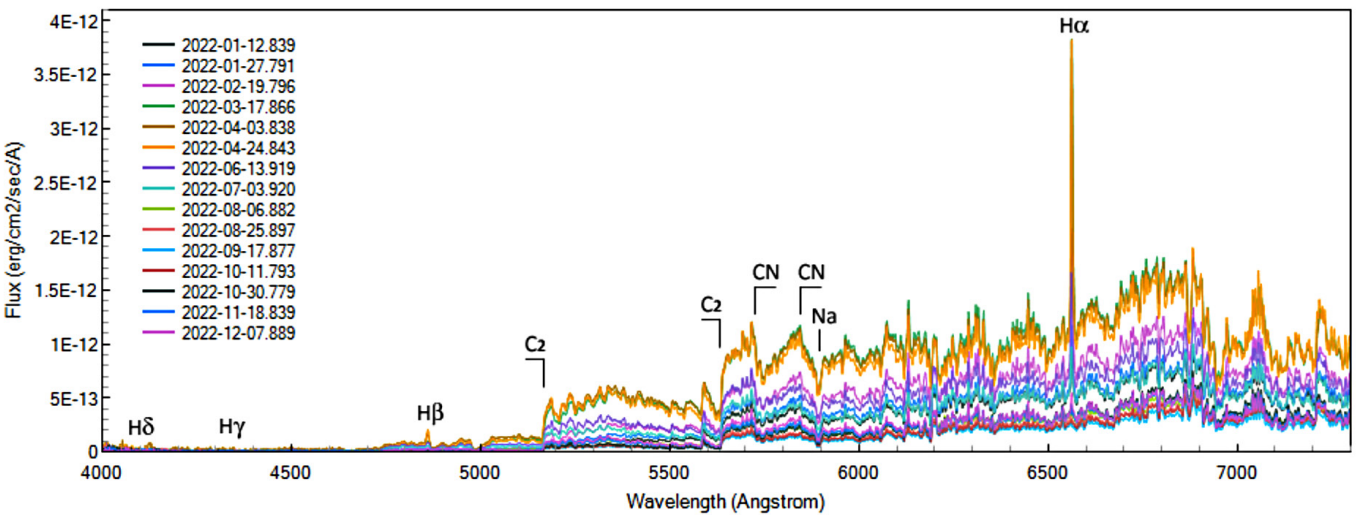

Figure 6. All flux calibrated spectra of OR Cep during the 2022 pulsation cycle marking the positions of the first 4 Balmer emission lines, the C2, and CN molecular band heads and the Na D absorption lines.

We have adopted an approach based on comparison with previously classified stars. In the case of OR Cep, we assigned a spectral type to each star using spectra in the Spectral Atlas of Carbon Stars in Barnbaum *et al.* (1996), which are classified according to the revised MK system in Keenan (1993). We downloaded spectra from Vizier (Barnbaum *et al.* 1996) for two stars in each of the spectral type categories A to D in Table 3 of Barnbaum *et al.* (1996) and merged the blue and red parts to form a single spectrum of each star. To find the closest match between each of our spectra of OR Cep with spectra from the Barnbaum Spectral Atlas, we first interpolated all spectra to a 1 Å grid, normalized them to a mean flux value of unity in the wavelength range 5610 Å to 5630 Å, and removed the regions around Balmer lines. We then found the spectrum from the Barnbaum Atlas which gave the least squared difference in flux with each of our OR Cep spectra integrated over the wavelength interval 4750 Å to 6500 Å. The spectra which best matched each of our OR Cep spectra, along with their spectral type from the Barnbaum Atlas, are listed in Table 4. During the brighter part of the pulsation cycle, our spectra of OR Cep best match spectral type C-N5+, while during the fainter part of the cycle they predominantly match spectral type C-J4.5. Although somewhat empirical, we feel this attempt at classification is the best we can achieve.

**4. Flux measurement**

As described earlier, AGB stars often experience pulsations which generate emission lines of the Balmer series during the peak of the pulsation cycle. To measure the absolute flux in these emission lines, we must subtract the contribution to this flux from the underlying local continuum. We did this by averaging the continuum flux in a 5 Å-wide region on either side of the line and interpolating the flux level under the line between those regions. We then subtracted this interpolated flux from the spectral profile of the line to obtain the flux emitted in the line. We estimated the uncertainty in line flux by combining the uncertainties in measuring the spectral profile of the line and in estimating the interpolated flux level under the line. Absolute flux in the first four Balmer lines for V667 Cas is listed in Table 5. Estimated uncertainties in line flux are ±11% for Hα, ±12% for Hβ, ±14% for Hγ, and ±24% for Hδ. Balmer line flux for OR Cep is listed in Table 6. Estimated uncertainties in line flux are ±13% for Hα, ±13% for Hβ, ±41% for Hγ, and ±49% for Hδ.

Figures 7 and 8 are composite plots showing how the absolute flux in Balmer emission lines in V667 Cas and OR Cep varied during the 2022 pulsation cycle. In V667 Cas there is clearly a progressive Balmer decrement from Hα to Hδ. In OR Cep, while the Hα emission line is of similar strength to that in V667 Cas, the Hβ line is much weaker, while both the Hγ and Hδ emission lines are barely undetectable.

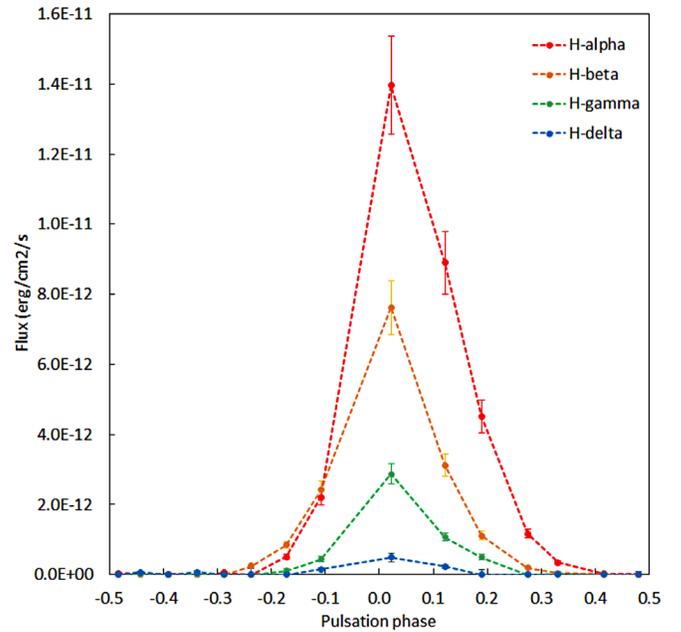

Figure 7. Composite plot of Balmer emission lines in V667 Cas showing their relative scale in flux and relationship in phase.

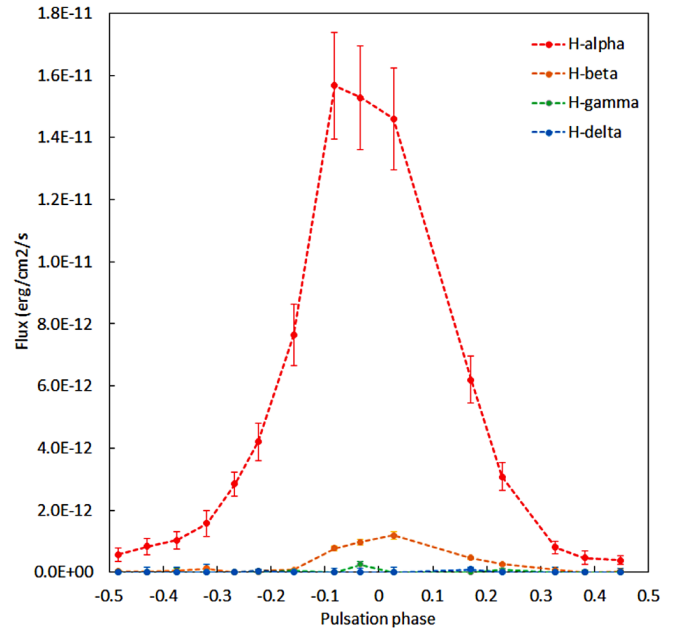

Figure 8. Composite plot of Balmer emission lines in OR Cep showing their relative scale in flux and relationship in phase.

Table 4. Standard spectra which best match each of our OR Cep spectra.

| JD | Phase | Best match standard spectrum | Spectral type |
|---|---|---|---|
| 2459592 | –0.27 | HD19881 | C–N5+ |
| 2459607 | –0.22 | HD19881 | C–N5+ |
| 2459630 | –0.16 | HD19881 | C–N5+ |
| 2459656 | –0.08 | HD19881 | C–N5+ |
| 2459673 | –0.03 | HD19881 | C–N5+ |
| 2459694 | 0.03 | HD19881 | C–N5+ |
| 2459744 | 0.17 | HD19881 | C–N5+ |
| 2459764 | 0.23 | HD19881 | C–N5+ |
| 2459798 | 0.33 | HO Cas | C–J4.5 |
| 2459817 | 0.38 | VX And | C–J4.5 |
| 2459840 | 0.45 | VX And | C–J4.5 |
| 2459864 | –0.48 | VX And | C–J4.5 |
| 2459883 | –0.43 | TW Oph | C–N5 |
| 2459902 | –0.37 | VX And | C–J4.5 |
| 2459921 | –0.32 | VX And | C–J4.5 |

Table 5. Julian Date, phase and Balmer line absolute flux for V667 Cas spectra recorded during the 2022 pulsation cycle.

| Julian Date | Phase | Hα line flux (ergs/cm2/sec) | Hβ line flux (ergs/cm2/sec) | Hγ line flux (ergs/cm2/sec) | Hδ line flux (ergs/cm2/sec) |
|---|---|---|---|---|---|
| 2459593 | –0.48 | 4.19E–14 | 0.00E+00 | 0.00E+00 | 0.00E+00 |
| 2459607 | –0.44 | 2.67E–14 | 0.00E+00 | 0.00E+00 | 6.84E–14 |
| 2459625 | –0.39 | 9.25E–15 | 0.00E+00 | 0.00E+00 | 0.00E+00 |
| 2459643 | –0.34 | 0.00E+00 | 0.00E+00 | 0.00E+00 | 6.42E–14 |
| 2459660 | –0.29 | 5.04E–14 | 0.00E+00 | 0.00E+00 | 0.00E+00 |
| 2459677 | –0.24 | 0.00E+00 | 2.50E–13 | 0.00E+00 | 0.00E+00 |
| 2459699 | –0.17 | 5.09E–13 | 8.51E–13 | 1.09E–13 | 0.00E+00 |
| 2459721 | –0.11 | 2.21E–12 | 2.42E–12 | 4.43E–13 | 1.54E–13 |
| 2459765 | –0.01 | 1.40E–11 | 7.63E–12 | 2.87E–12 | 4.95E–13 |
| 2459799 | 0.02 | 8.90E–12 | 3.12E–12 | 1.08E–12 | 2.36E–13 |
| 2459822 | 0.12 | 4.52E–12 | 1.12E–12 | 5.01E–13 | 0.00E+00 |
| 2459851 | 0.19 | 1.17E–12 | 1.95E–13 | 0.00E+00 | 0.00E+00 |
| 2459870 | 0.19 | 3.48E–13 | 4.61E–14 | 2.50E–14 | 0.00E+00 |
| 2459899 | 0.27 | 2.83E–14 | 2.19E–14 | 0.00E+00 | 0.00E+00 |
| 2459921 | 0.33 | 1.33E–14 | 0.00E+00 | 0.00E+00 | 0.00E+00 |

Notes: Line flux too small to measure is shown as zero. Estimated uncertainty in line flux is ±11% for Hα, ±12% for Hβ, ±14% for Hγ, and ±24% for Hδ.

Table 6. Julian Date, phase and Balmer line absolute flux for OR Cep spectra recorded during the 2022 pulsation cycle.

| Julian Date | Phase | Hα line flux (ergs/cm2/sec) | Hβ line flux (ergs/cm2/sec) | Hγ line flux (ergs/cm2/sec) | Hδ line flux (ergs/cm2/sec) |
|---|---|---|---|---|---|
| 2459592 | –0.27 | 2.84E–12 | 6.33E–15 | 0.00E+00 | 0.00E+00 |
| 2459607 | –0.22 | 4.20E–12 | 5.49E–14 | 0.00E+00 | 4.76E–14 |
| 2459630 | –0.16 | 7.64E–12 | 7.15E–14 | 5.30E–14 | 0.00E+00 |
| 2459656 | –0.08 | 1.57E–11 | 7.73E–13 | 0.00E+00 | 0.00E+00 |
| 2459673 | –0.03 | 1.53E–11 | 9.71E–13 | 2.48E–13 | 0.00E+00 |
| 2459694 | 0.03 | 1.46E–11 | 1.19E–12 | 0.00E+00 | 0.00E+00 |
| 2459744 | 0.17 | 6.20E–12 | 4.58E–13 | 0.00E+00 | 1.01E–13 |
| 2459764 | 0.23 | 3.08E–12 | 2.66E–13 | 8.35E–14 | 0.00E+00 |
| 2459798 | 0.33 | 8.00E–13 | 7.69E–14 | 0.00E+00 | 0.00E+00 |
| 2459817 | 0.38 | 4.70E–13 | 0.00E+00 | 0.00E+00 | 0.00E+00 |
| 2459840 | 0.45 | 3.91E–13 | 3.91E–14 | 0.00E+00 | 0.00E+00 |
| 2459864 | –0.48 | 5.66E–13 | 3.35E–14 | 0.00E+00 | 0.00E+00 |
| 2459883 | –0.43 | 8.28E–13 | 2.23E–14 | 0.00E+00 | 0.00E+00 |
| 2459902 | –0.37 | 1.04E–12 | 4.98E–14 | 0.00E+00 | 0.00E+00 |
| 2459921 | –0.32 | 1.58E–12 | 1.20E–13 | 0.00E+00 | 0.00E+00 |

Notes: Line flux too small to measure is shown as zero. Estimated uncertainty in line flux is ±13% for Hα, ±13% for Hβ, ±41% for Hγ, and ±49% for Hδ.

## 5. Summary

We observed the 2022 pulsation cycles of the S-type Mira V667 Cas and the carbon star Mira OR Cep using concurrent spectroscopy and photometry. We measured how their V and $I_c$ magnitudes and V-$I_c$ color index varied over the cycle. We classified their spectral types according to criteria of the revised MK system and measured and compared how the absolute flux emitted by the hydrogen Balmer lines in these stars varied during the cycle.

## 6. Acknowledgements

We are grateful to the anonymous referee for a helpful review which has improved the paper and to Lee Anne Willson for her advice and encouragement to observe these stars. This research made use of the AAVSO Photometric All-Sky Survey (Henden 2021). We gratefully acknowledge the work of developers of the Astropy package and other contributors to this valuable community resource.